\documentclass
[preprint,showpacs,showkeys,byrevtex,superscriptaddress,prc]{revtex4}%
\usepackage{amssymb}
\usepackage{amsmath}
\usepackage{graphicx}
\usepackage{amsfonts}%
\ifx\pdfoutput\relax\let\pdfoutput=\undefined\fi
\newcount\msipdfoutput
\ifx\pdfoutput\undefined\else
\ifcase\pdfoutput\else
\ifx\paperwidth\undefined\else
\ifdim\paperheight=0pt\relax\else\pdfpageheight\paperheight\fi
\ifdim\paperwidth=0pt\relax\else\pdfpagewidth\paperwidth\fi
\fi\fi\fi
\begin{document}
\title{New determination of proton spectroscopic factors and reduced widths for
$^{8}$Be states in the $16.5-18.0$ MeV excitation energy region via the study
of the $^{7}$Li($^{3}$He,d)$^{8}$Be transfer reaction at E$_{lab}$ $=\ 20$
MeV. }
\author{A. Belhout}
\affiliation{Universit\'{e} des Sciences et de la Technologie Houari Boumediene (USTHB),
Facult\'{e} de Physique, BP 32, El-Alia, 16111 Bab Ezzouar, Algiers, Algeria,}
\author{S. Ouichaoui$^{\ast}$}
\affiliation{Universit\'{e} des Sciences et de la Technologie Houari Boumediene (USTHB),
Facult\'{e} de Physique, BP 32, El-Alia, 16111 Bab Ezzouar, Algiers, Algeria,}
\author{H. Beaumevieille}
\affiliation{Universit\'{e} des Sciences et de la Technologie Houari Boumediene (USTHB),
Facult\'{e} de Physique, BP 32, El-Alia, 16111 Bab Ezzouar, Algiers, Algeria,}
\author{A. Bouchemha}
\affiliation{Universit\'{e} des Sciences et de la Technologie Houari Boumediene (USTHB),
Facult\'{e} de Physique, BP 32, El-Alia, 16111 Bab Ezzouar, Algiers, Algeria,}
\author{G. Bogaert}
\affiliation{Centre de Sciences Nucl\'{e}aires et des Sciences de la Mati\`{e}re (CSNSM),
IN2P3/CNRS-Universit\'{e} de Paris XI, 91405 Campus Orsay France,}
\author{S. Fortier}
\affiliation{Institut de Physique Nucl\'{e}aire, IN2P3/CNRS-Universit\'{e} de Paris XI,
91406 Orsay Cedex, France,}
\author{J. Kiener}
\affiliation{Centre de Sciences Nucl\'{e}aires et des Sciences de la Mati\`{e}re (CSNSM),
IN2P3/CNRS-Universit\'{e} de Paris XI, 91405 Campus Orsay France,}
\author{A. Lefebvre-Schuhl}
\affiliation{Centre de Sciences Nucl\'{e}aires et des Sciences de la Mati\`{e}re (CSNSM),
IN2P3/CNRS-Universit\'{e} de Paris XI, 91405 Campus Orsay France,}
\author{J. M. Maison}
\affiliation{Institut de Physique Nucl\'{e}aire, IN2P3/CNRS-Universit\'{e} de Paris XI,
91406 Orsay Cedex, France,}
\author{L. Rosier}
\affiliation{Institut de Physique Nucl\'{e}aire, IN2P3/CNRS-Universit\'{e} de Paris XI,
91406 Orsay Cedex, France,}
\author{J. Rotbard}
\affiliation{Institut de Physique Nucl\'{e}aire, IN2P3/CNRS-Universit\'{e} de Paris XI,
91406 Orsay Cedex, France,}
\author{V. Tatischeff}
\affiliation{Centre de Sciences Nucl\'{e}aires et des Sciences de la Mati\`{e}re (CSNSM),
IN2P3/CNRS-Universit\'{e} de Paris XI, 91405 Campus Orsay France,}
\author{J.P. Thibaud}
\affiliation{Centre de Sciences Nucl\'{e}aires et des Sciences de la Mati\`{e}re (CSNSM),
IN2P3/CNRS-Universit\'{e} de Paris XI, 91405 Campus Orsay France,}
\author{J. Vernotte}
\affiliation{Institut de Physique Nucl\'{e}aire, IN2P3/CNRS-Universit\'{e} de Paris XI,
91406 Orsay Cedex, France,}

\begin{abstract}
The angular distributions of $^{8}$Be states in the excitation energy region,
\textit{E}$_{x}$ $\ \sim$($16.5-18.2$) MeV, produced in the $^{7}$Li($^{3}%
$He,d)$^{8}$Be proton transfer reaction have been measured at the Orsay
14.8$-$MV tandem accelerator for $^{3}$He$^{2+}$ ion bombarding energy,
\textit{E}$_{lab}$ $=20$ MeV, and forward anglular range, $\theta_{lab}=5^{%
{{}^\circ}%
}-50%
{{}^\circ}%
$. A high energy resolution detection system composed of a split-pole magnetic
spectrometer and a $\Delta E-E$,\ position-sensitive drift chamber was used to
record the energy spectra of outgoing deuterons. The measured cross section
data for the direct reaction component have been separated from the compound
nucleus one, then analyzed in the framework of the non local, FR-DWBA theory.
New values of the $C^{2}S$ and ($S_{p1/2},$ $S_{p3/2})$ proton absolute and
partial spectroscopic factors and related $\gamma_{p}^{2}(a)$\ proton reduced
widths versus the p $+$ $^{7}$Li channel radius have been extracted for the
$2^{+}(16.626)$ and $2^{+}(16.922),T=0+1$ isospin-mixed loosely bound states
of astrophysical interest and the $1^{+}(17.640)$, $T=1$ unbound state of
$^{8}$Be. They are compared to sparse earlier experimental values and to
shell-model predicted ones from the literature, and are discussed. In
particular, the status of the spectroscopic information on the $2^{+}$
isospin-mixed doublet is reviewed and up-dated. The application in nuclear
astrophysics of the DWBA derived results is emphasised.

\end{abstract}
\preprint{HEP/123-qed}
\date{July 2017}

\email{$^{\ast}$Corresponding author: \textit{souichaoui@gmail.com.} }
\maketitle
\tableofcontents
\volumeyear{year}
\volumenumber{number}
\issuenumber{number}
\eid{identifier}
\received[Received text]{date}

\revised[Revised text]{date}

\accepted[Accepted text]{date}

\published[Published text]{date}

\startpage{1}
\endpage{ }

\section{Introduction}

The spectroscopic information on nuclear energy levels (excitation energies,
spectroscopic factors, particle reduced widths,..) is crucial in nuclear
physics regarding the structure of light and heavier nuclei. It is intensively
used for describing, modeling and elucidating many nuclear structure problems
such as isobaric analogue states, rotational bands, isospin mixing, two-level
systems, $\alpha$-particle clustering or exotic nuclear states (see
\cite{1,2,3,4,5} and references therein). This information is also of great
interest to nuclear astrophysics where nuclear levels of particular structure
often play a prominent role in big bang and/or in stellar nucleosyntheses
(BBN, SN) \cite{6,7,8}. This is the case, for example, of the $J^{\pi}=2^{+},$
$T=1$ ground states of $^{8}$Li and $^{8}$B nuclei and the two $2^{+}$,
$T=0+1$ isospin-mixed states in the two $\alpha-$cluster $^{8}$Be nucleus at
E$_{x}$ = $16.626$ MeV and $16.922$ MeV below the $p$ $+$ $^{7}Li$\ threshold
(E$_{x}$ $=$ $17.255$ MeV) \cite{9,10,11,12,13,14,15}. In particular, the
latter two loosely bound states are involved in\ the resonant $^{7}%
$Li(p,$\alpha$)$^{4}$He hydrogen burning reaction implied both in BBN and in
SN, whose cross section at stellar energies is small ($\sigma$ $=$
$(4.3\pm0.9)\times10^{-5}$ mb at \textit{E}$_{p}$ = $28.1$ keV)\ and difficult
to measure directly due mainly to the inhibitory effect of the $p$ $+$
$^{7}Li$ Coulomb barrier ($B_{c}=2.473$ MeV). Alternatively, the spectroscopic
information on the latter two $2^{+}$ states, notably their proton reduced
widths, should be much more easily accessible via the $^{7}$Li($^{3}%
$He,d)$^{8}$Be transfer reaction. Indeed, the cross section of the latter
reaction is large and the properties of $^{8}$Be states can be more easily
derived from it provided the energy of the incident $^{3}$He$^{2+}$ ion beam
is sufficiently high for the direct interaction mechanism to be dominating
over the ($^{10}$B) compound nucleus formation. Then, the astrophysical $S(E)$
$-$ factor and the stellar rate of the $^{7}$Li(p,$\alpha$)$^{4}$He proton
capture reaction may be efficiently determined indirectly via the measurement
of$\ $the$^{7}$Li($^{3}$He,d)$^{8}$Be transfer reaction angular distributions.
Besides, other indirect methods (such as the asymptotic normalization
coefficient (ANC), the trojan-horse (TH) and the Coulomb break-up methods
\cite{16}) can be also used to reach the same objectives. Furthermore
regarding astrophysical applications, the most precise possible determination
of the rates of nuclear reactions involving the $^{6,7}$Li and $^{7,8}$Be
isotopes and considered in BBN calculations have been recommended (see, e.g.,
\cite{17} and references therein). The objective in sight was to study the
origin of discrepancies between the observed abundance of $^{7}$Li in
metal-poor galactic halo dwarf stars and its predicted primordial abundance
($^{7}$Li/H $=$ ($1.58\pm0.314$) $\times$ $10^{-10}$ \cite{18} and
($4.68\pm0.67$) $\times10^{-10}$ \cite{19}, respectively), and between an
assumed to be observed value \cite{20} of the Lithium isotopic ratio and its
predicted one ($^{6}$Li/$^{7}$Li $\sim10^{-5}$ \cite{21}). However, an
experiment of the LUNA collaboration \cite{22} constraining the $^{2}%
$H($\alpha,\gamma)^{6}$Li reaction cross section yieded a value of this ratio,
$^{6}$Li/$^{7}$Li $=$ ($1.5\pm0.3$) $\times10^{-5}$, matching the BBN
prediction \cite{21}, which has been also confirmed recently by theoretical
calculations \cite{23}. Then, as highlighted by Coc in \cite{24}, while a BBN
$^{6}$Li problem is no longer up-to-date, only the $^{7}$Li problem still
persists presently. But among other hardly searched for solutions (see
\cite{24} and references therein), the perspective of solving the latter
puzzle via nuclear reaction rate evaluations seems not to hold any more
actually \cite{24}. Note, besides, that following the recent observation of a
6.8$\sigma$ anomaly at E$_{x}$ $\sim$ 17 MeV in $^{\emph{8}}$Be \cite{25}
decaying via internal electron-positron pair creation, further perspectives
for a particle physics solution to the cosmological Lithium problem have been
proposed (see \cite{26} and references therein).

In one-step transfer reactions such as the $^{7}$Li($^{3}$He,d)$^{8}$Be one -
e.g., proton stripping in (d,n) and ($^{3}$He,d) reactions or $\alpha
$-particle stripping in ($^{6}$Li,d) and ($^{7}$Li,t) reactions - one particle
is selectively transferred from the projectile into a given shell of the final
nucleus with definite $n\ell j$ quantum numbers without altering the target
nucleus core \cite{27}. Such direct reactions have been used since long as
privileged tools in order to precisely determine the level parameters
(excitation energies, widths, J$^{\pi}$ values) for many involved residual
nuclei. The accumulated experimental data on the energy levels of the $A$
$=8-10$ light nuclei has been reported in the successive compilations by
Ajzenberg-Selove and Tilley et al. (see \cite{28} and references therein).
Many nuclear reactions and various experimental methods have been used for
determining the parameters of the peculiar $2^{+}$ isospin-mixed\ doublet of
$^{8}$Be. However, inconsistencies between the results from different groups
have been observed due evenly to the complexity of the investigated nuclear
interaction processes involving interference effects. Thus, the spectroscopic
information on $^{8}$Be states in the excitation energy region, E$_{x}%
\sim(16.5-18)$ MeV, is yet lacking in the literature and some nuclear
reactions involving the $2^{+}$ isospin-mixed doublet have not been
sufficiently explored. Especially, only few measurements have been carried out
previously \cite{10,29,30,31} on the $^{7}$Li($^{3}$He,d)$^{8}$Be transfer
reaction with cross section experimental data being reported only in reference
\cite{30}. Indeed, Marion et al. \cite{10} have measured a unique deuteron
energy spectrum at a laboratory angle, $\theta_{lab}=40%
{{}^\circ}%
$, for $^{3}$He ion bombarding energy, \textit{E}$_{lab}$ \ $=$ $10.972$ MeV,
in order to determine the total widths of the $2^{+}(16.626)$ and
$2^{+}(16.922)$\ states of $^{8}$Be. The study of this reaction by Piluso at
al. \cite{29} was limited to recording only two deuteron energy spectra at two
laboratory angles, $\theta_{lab}=10%
{{}^\circ}%
$ and $25%
{{}^\circ}%
$, for a $^{3}$He ion bombarding energy, \textit{E}$_{lab}=15$ MeV, which were
used to determine level parameters for the above two $2^{+}$ states of $^{8}%
$Be with pointing out their interference contributions and to search for
$1$p$-1$h states. Besides, Basak et al. \cite{30} have measured the $^{7}%
$Li($\overrightarrow{^{3}He}$,d)$^{8}$Be reaction angular distributions for
polarized $^{3}$He ions of incident energy, \textit{E}$_{lab}$ $=33.3$ MeV.
The analysis of the latter data within the DWBA formalism has led these
authors to derive the only available $C^{2}S_{global}$\ proton spectroscopic
factor experimental values for the $2^{+}$($16.626$), $1^{+}$($17.64)$ and
$1^{+}(18.15)$ states of $^{8}$Be from this reaction, to our knowledge, while
the $2^{+}$($16.922)$ state seemed to be not clearly populated in that
experiment. Finally, Cocke has reported \cite{31} a global experimental
angular distribution for the $2^{+}$($16.626)$ state measured at
\textit{E}$_{lab}$ $=10$ MeV where the direct and compound nucleus
contributions were not separated. Then, despite many efforts devoted to study
the $2^{+}$, $T=0+1$ isospin-mixed doublet of $^{8}$Be (see
\cite{9,10,11,12,28,29,30,31} and references therein), not only the
corresponding spectroscopic information remains incomplete but the structures
of these two special states steadily seem to be complex and not well
elucidated. On the other hand concerning the reaction mechanism prevailing at
thermonuclear temperatures in the $^{7}$Li(p,$\alpha$)$^{4}$He proton capture
reaction, a previous DWBA analysis \cite{32} of precise cross section
experimental data available for\textit{ E}$_{p}$ $=$ ($13$ $-10^{3}$) keV
proton energies \cite{32,33} have led the authors to an apparently good
agreement between theory and experiment, thus suggesting to interpret the data
in terms of a dominant three-nucleon direct reaction transfer. However, this
conclusion contradicts existing clear evidences for $^{8}$Be compound nucleus
formation in this reaction, as will be detailed later in Section III.
Therefore, the $^{7}$Li(p,$\alpha$)$^{4}$He reaction cross section
experimental data can be more pertinently analysed, instead, in the framework
of the R-Matrix theory with assuming the predominance of ($^{8}$Be) compound
nucleus reaction mechanism. In particular, the $\gamma_{p}^{2}$ ($a$)\ proton
reduced widths for $^{8}$Be states (mainly the $2^{+}$ isospin-mixed doublet)
derived as free fit parameters in such analysis deserve to be compared to
experimental counterparts from the$^{7}$Li($^{3}$He,d)$^{4}$He or $^{7}%
$Li(d,n)$^{8}$Be proton stripping reactions and to shell model predictions
\cite{34}. It therefore appeared to us worthwhile to critically re-visit the
spectroscopy of $^{8}$Be nucleus states within the $(16.5-18.2)$ MeV
excitation energy region including the $2^{+},$ $T=0+1$ isospin mixed doublet.
For all these reasons, we have undertaken the measurement of the $^{7}%
$Li($^{3}$He,d)$^{8}$Be reaction angular distributions for $^{3}$He$^{2+}$ ion
bombarding energy, $E_{lab}=20$ MeV. In this work, we thus preferentially
aimed at performing a new and reliable experimental determination of the
$C^{2}S$ spectroscopic factors for these two $2^{+}$ states of astrophysical
interest and deriving relevant values of the closely related proton reduced
widths, $\gamma_{p}^{2}$, that can be very useful, e.g., for constraining the
number of fit parameters in the R-Matrix analyses of cross section
experimental data for fusion reactions involving $^{8}$Be states.

A high energy resolution, position-sensitive detection system was used in the
experiment that will be described in Section II. The contents of the direct
interaction component within the recorded energy spectra of the outgoing
deuterons have been separated from the compound nucleus contribution and
transformed into corresponding center of mass cross sections. Then, the latter
angular distribution experimental data have been carefully analyzed in the
framework of the non-local, finite-range DWBA formalism. A detailed account of
the performed theoretical analysis of the $^{7}$Li($^{3}$He,d)$^{8}$Be angular
distribution experimental data is given in Section III where the derived
$C^{2}S$ and $\gamma_{p}^{2}$\ data for the three excited states of $^{8}$Be
considered here, i.e., the $2^{+}(16.626)$, $2^{+}$($16.922)$ and
$1^{+}(17.64)$ states, are reported and discussed. Finally, a summary and
conclusions are given in Section IV.

\section{Experimental details, procedures, results and discussions}

\subsection{Experimental set up and procedure}

The experimental set up and detection system were the same as in our previous
experiment on the $^{12}$C($^{6}$Li,d)$^{16}$O $\alpha$-transfer reaction
\cite{35}. Then, mainly specific aspects to the current $^{7}$Li($^{3}%
$He,d)$^{8}$Be experiment are emphasized in this section.

The experiment was carried out at the Orsay-Institut de Physique Nucl\'{e}aire
MP Tandem accelerator, on the line of the\ Enge split-pole magnetic
spectrometer \cite{36}. A 20 MeV $^{3}$He$^{2+}$ ion beam delivered with high
energy resolution ($\Delta E/E\approx2\times10^{-4}$) and average beam current
intensity of $\sim100$ $nA$ was directed onto a self-supporting, $49$
$\mu$%
g.cm$^{-2}$-thick target foil of natural lithium placed under high vacuum in
the reaction chamber at the object focal point of the magnetic
spectrometer.\textbf{ }The target was prepared by vacuum evaporation of
metallic lithium. Before being used in the experiment, it was continuously
maintained under vacuum in order to reduce its oxidation and/or contamination
by chemical impurities until it was introduced into the reaction chamber by
means of a sieve without breaking the vacuum. The nuclear reaction products
were, first, momentum analyzed by the magnetic spectrometer whose horizontal
entrance aperture was set at $\pm$ $1.5%
{{}^\circ}%
$ corresponding to a solid angle, $\Delta\Omega,$ of $\sim$ $1.6$ msr. Then,
they were identified in the spectrometer image focal plane by a detection
system of $70$ cm length composed of three successive detectors \cite{37}: (i)
a position sensitive $128-$ anode wires drift chamber giving the position,
$X$, of particle impacts, (ii) a proportional counter measuring the particle
energy loss, $\Delta E$, and (iii) a plastic scintillator (associated to a
photomultiplier tube through a light-guide) measuring the particle residual
energy, $E^{\prime}=E-\Delta E$. Both the target thickness and the beam
current intensity (the latter was measured by a well shielded and isolated
Faraday cup) were continuously monitored during the whole experiment by means
of a $100$ $%
\mu
m-$ thick surface barrier Si detector placed inside the reaction chamber at
$\theta_{lab}=42%
{{}^\circ}%
$ relative to the incident beam direction. The $^{7}$Li($^{3}$He,d)$^{8}$Be
reaction angular distributions were measured by recording energy spectra of
the outgoing deuterons over the forward angular range, $5%
{{}^\circ}%
\leq\theta_{lab}\leq50%
{{}^\circ}%
$, in $5%
{{}^\circ}%
$ steps.

\subsection{Target thickness determination}

The thickness of the used Li target has been determined as follows. The energy
spectra for $25$ MeV $^{3}$He$^{++}$ ions elastically scattered off a LiF
target were registered at observation angles, $\theta_{lab}=36%
{{}^\circ}%
,39%
{{}^\circ}%
,42%
{{}^\circ}%
,45%
{{}^\circ}%
$ and $48%
{{}^\circ}%
$ with the magnetic spectrometer being set to focus onto the detector the
$^{3}$He particles scattered off $^{19}$F nuclei. The number of these nuclei
in the LiF target, $N(^{19}F)=(35.07\pm01.46)\times10^{17}cm^{-2}$, was
derived from the $^{19}$F($^{3}$He,$^{3}$He)$^{19}$F elastic scattering cross
sections measured previously \cite{38} at the same ion energy and laboratory
angles. Consideration of the energy spectra from the monitor detector and of
the accumulated beam charge showed that the target remained stable during
these elastic scattering measurements. Assuming the conservation of the
stoichiometric ratio during the fabrication of the LiF target leads to
$N(^{19}F)=N(Li)$. Then, setting the spectrometer such that the $^{3}$He
particles scattered off Li nuclei were focused onto the detector, two spectra
for elastically scattered $25$ MeV $^{3}$He$^{++}$ ions were registered at
$\theta_{lab}$ $=36%
{{}^\circ}%
$: one with the LiF target in place, the other with the metallic Li target.
Then, the comparison of the areas of the elastic scattering peaks leads to the
number of Li nuclei contained in the metallic Li target, $N(Li)=(42.80\pm
02.00)\times10^{17}cm^{-2}$, i.e., to a thickness of $\sim$ $49\pm3$ $%
\mu
g$ $cm^{-2}.$ One then deduces the number of $^{7}$Li nuclei to be
$N(^{7}Li)=(39.64\pm01.84)\times10^{17}cm^{-2}$. Note that the ratio of the
number of Li nuclei in the metallic Li target to those in the LiF target,
$N(Li)_{Li\text{ }target}/N(Li)_{LiF\text{ }target}$, amounts to $1.22\pm
0.08$, this value being obtained without taking into account the information
from the Si monitor detector. It is in very good agreement with the ratio,
$A(Li)_{Li\text{ }target}/A(Li)_{LiF\text{ }target}=1.21\pm0.02$, with A
denoting the ratio of the number of counts of Li nuclei to the accumulated
beam charge, Q (corrected for the dead-time), from the monitor detector, i.e.,
$A=(N_{counts}^{\prime}(Li)/Q)_{Monitor}$.

\subsection{Analysis of the deuteron energy spectra}

Despite the precautions taken during the fabrication and handling of the Li
target, the deuteron energy spectra exhibited significant contamination in
$^{12}$C, $^{14}$N, $^{16}$O and $^{19}$F. Then, peaks corresponding to
($^{3}$He, d) proton stripping reactions on the corresponding nuclei giving,
respectively, $^{13}$N, $^{15}$O, $^{17}$F and $^{20}$Ne as residual nuclei
have been observed. Part of the deuteron energy spectrum taken at
$\theta_{lab}=5%
{{}^\circ}%
$ and showing deuteron peaks corresponding to the states of $^{8}$Be at
$E_{x}$ $=16.626$ MeV ($2^{+}$), $16.922$ MeV ($2^{+}$), $17.640$ MeV ($1^{+}%
$) and $18.150$ MeV ($1^{+}$) is reported in Fig. 1. Note, however, that owing
to the fact that the peak associated with the relatively broad ($\Gamma
_{cm}=138\pm6$ keV \cite{28}) $1^{+}$($18.150$), $T=1$ state was more
critically affected by contaminant peaks from secondary reactions, this state
was not considered in the current study. In all the measured deuteron energy
spectra, the peaks associated with the two $2^{+}$ ($16.626$) and $2^{+}%
$($16.922$) states of main concern here were quite well separated due to the
high energy resolution of the detection system used, with the peak for the
former state being considerably more intense than the one for the latter state
(see also Fig. 2).\textbf{ }Furthermore, consistently with kinematics
predictions, these two peaks did not suffer severe contamination, notably from
the $^{16}$O($^{3}$He,d)$^{17}$F reaction beyond\textbf{ }$\theta_{lab}\sim15%
{{}^\circ}%
$\textbf{.}\textit{ }While the contents of the peaks associated with the
narrow $1^{+}$($17.640$), ($T=1$) state ($\Gamma_{cm}=10.7\pm\ 0.5$ keV
\cite{10,28}) were extracted easily, those of the peaks for the $2^{+}%
$($16.626$) and $2^{+}$($16.922$) states required a special treatment. Indeed,
as indicated, these two final states are both characterized by an important
$T$ $=0+1$ isospin mixing \cite{9,10,11}. Then, their respective wave
functions can be written as (see \cite{10})%
\begin{equation}
\left\vert 16.626\text{ }MeV\right\rangle =A\left\vert T=0\right\rangle
+B\left\vert T=1\right\rangle \label{eq.1}%
\end{equation}%
\begin{equation}
\left\vert 16.922\text{ }MeV\right\rangle =B\left\vert T=0\right\rangle
-A\left\vert T=1\right\rangle \label{eq.2}%
\end{equation}
where the coefficients (A, B) represent respective weights of the $\left\vert
T=0\right\rangle $ and $\left\vert T=1\right\rangle $ isospin contributions,
expected to have close values and to fulfill the normalization condition%
\begin{equation}
A^{2}+B^{2}=1. \label{eq.3}%
\end{equation}
Considering that these two $2^{+}$ states are simultaneously populated by
direct and compound nucleus reaction mechanisms, we have developed formula
($28$) from reference \cite{10}, thus obtaining the following expression for
the corresponding center of mass differential cross section.

\begin{multline}
\frac{d^{2}\sigma}{d\Omega dE}(\theta)=\frac{N_{d}^{2}(\theta)(B+A)^{2}%
+N_{c}^{2}(\theta)(C^{2}+D^{2})}{(E-E_{1})^{2}+\Gamma_{1}^{2}/4}+\frac
{N_{d}^{2}(\theta)(B-A)^{2}+N_{c}^{2}(\theta)(C^{2}+D^{2})}{(E-E_{2}%
)^{2}+\Gamma_{2}^{2}/4}\label{eq.4}\\
+2N_{d}^{2}(\theta)(B+A)(B-A)\frac{(E-E_{1})(E-E_{2})+\Gamma_{1}\Gamma_{2}%
/4}{\left[  (E-E_{1})^{2}+\Gamma_{1}^{2}/4\right]  \left[  (E-E_{2}%
)^{2}+\Gamma_{2}^{2}/4\right]  }.
\end{multline}

In this expression, indices $1$ and $2$ refer, respectively, to the properties
of the $2^{+}$ $(16.626)$ and $2^{+}$ ($16.922$) states, the angle-dependent
parts for the direct ($d$) and compound nucleus ($c$) contributions are
contained, respectively, in the $N_{d}(\theta)$ and $N_{c}(\theta)$ terms, and
C and D are the amplitudes of the compound nucleus states. Equation \ref{eq.4}
then involves two Breit-Wigner shapes describing the two $2^{+\text{ }}$states
and their interference term. The nuclear level parameters entering in it were
fitted using the CERN computer program PAW \cite{39} to reproduce the
experimental deuteron energy spectra. The fit parameters that had to be
considered in this analysis were: $E_{1}$, $E_{2}-E_{1}$, $\Gamma_{1}$,
$\Gamma_{2}$, $\alpha=N_{d\text{ }}(\theta)$ $(B+A)$, $\beta=N_{d\text{ }%
}(\theta)$ $(B-A)$ and$\ \gamma=N_{c}$ $(\theta)$ $(C^{2}+D^{2})$. Then, this
set of seven free physics parameters was further reduced to only four with
adopting well established c. m. values of the level widths and resonance
energies for the $2^{+}$\ isospin-mixed doublet from the literature
\cite{28}\emph{,} i.e., the $\Gamma_{1}$, $\Gamma_{2}$ and $E_{2}-E_{1}$
values reported in table I. The analysis of the deuteron energy spectra
carried out using equation \ref{eq.4} has led us to a satisfactory
reproduction of the shapes of the measured deuteron energy spectra. The
corresponding fit parameters for the extracted direct reaction contribution in
this equation are reported in table I (see the text below in this section).
The spectrum recorded at $\theta_{lab}=35%
{{}^\circ}%
$, showing the peaks associated with the two $2^{+}$ states of $^{8}$Be well
isolated and separated from each other, is reported in Fig. 2 where the solid
curve represents the best fit to the experimental data points generated by the
PAW software.\emph{ }As can be seen in this figure, the $2^{+}(16.626)$ state
is then much more populated than the $2^{+}(16.922)$ upper state in the $^{7}%
$Li($^{3}$He,d)$^{8}$Be proton stripping reaction at bombarding energy,
E$_{lab}=20$ MeV. Indeed, the ratio of \ the peak intensities (peak content
areas) for these states,$\frac{I_{16.626}}{I_{16.922}},$ amounts here to $5.3$
for $\theta_{lab}=35%
{{}^\circ}%
$ and increases with decreasing the observation angle (see Figs. (1, 2)). This
behavior is consistent with the measurements of Piluso et al. on this transfer
reaction at E$_{lab}=15$ MeV (see Fig. 3 of reference \cite{29}), while at the
lower $^{3}$He ion energy of $\sim11$ MeV the peaks associated with the two
$2^{+},$isospin-mixed states of $^{8}$Be were found to be of comparable
heights and sizes (see references \cite{10,31}).

The quantities ($A$, $B$, $\alpha$, $\beta$) associated with the direct
reaction component in equation \ref{eq.4} are related by%

\begin{equation}
\frac{A}{B}=\frac{\alpha-\beta}{\alpha+\beta}. \label{eq.5}%
\end{equation}
Using this relation and the normalization condition of equation \ref{eq.3},
one can deduce the values of coefficients $A$, $B$ and parameter $N_{d}$
$(\theta)$ for the direct reaction component. Then, values of the quantities
$A/B,$ $A+B$ and $A-B$ have been obtained in the analysis of the different
deuteron energy spectra recorded in this experiment. Averaging all $A$ values,
we derived the $\left\langle A\right\rangle $, $\left\langle B\right\rangle $
and $\left\langle A/B\right\rangle $ mean values listed in table I that show
to be in fairly good agreement with the $A$, $B$ and $A/B$ values obtained by
Marion et al. \cite{10} in their earlier study of the $^{7}$Li($^{3}%
$He,d)$^{8}$Be transfer reaction at E$_{lab}\thicksim$ $11$ MeV\textbf{.} It
therefore appears that the values of the ($A$, $B$) amplitudes associated with
the direct reaction component in equation \ref{eq.4} are independent on the
reaction bombarding energy.

Only parts of the peak areas of interest in the measured deuteron spectra due
to the direct reaction component have been considered in our non local,
FR-DWBA analysis (see Section III) of the measured $^{7}$Li($^{3}$He,d)$^{8}%
$Be angular distribution data. These peak areas were calculated using the relation%

\begin{equation}
Area_{i}=\frac{2\pi H_{i}^{2}}{\Gamma_{i}} \label{eq.6}%
\end{equation}
where $H_{i}=N_{d}$ $(\theta)$ $(A+B)$ or $N_{d}$ $(\theta)$ $(A-B)$ depending
on whether index i $=$ $1$ or $2$, with $N_{d}$ $(\theta)$ denoting the
angle-dependent part of the direct reaction component, as stated. Then, the
derived count numbers for this component have been transformed into
corresponding center of mass differential cross sections.

\begin{center}%
\begin{table}[tbp] \centering
%

\begin{tabular}
[c]{|c|c|c|c|c|c|c|}\hline
& $E_{2}-E_{1}$ (keV) & $\Gamma_{1}$ (keV) & $\Gamma_{2}$ (keV) &
$\left\langle A\right\rangle $ & $\left\langle B\right\rangle $ &
$\left\langle A/B\right\rangle $\\\hline
This work & $296\pm6$ & $108.1$ & $74.0$ & $0.797\pm0.036$ & $0.604\pm0.08$ &
$1.32\pm0.08$\\\hline
\emph{\cite{10}} & $274+3$ & $\ \ 113\pm3$ & $\ 73\pm3$ & $0.772$ & $\ 0.636$
& $1.47\pm0.07$\\\hline
\end{tabular}
%

\end{table}%

\end{center}

\subsection{Experimental angular distributions}

The derived angular distribution experimental data for the three states of
$^{8}$Be studied here (i.e., the $2^{+}$($16.626$), $2^{+}$($16.922$) and
$1^{+}$($17.64$) excited states) are reported in Figs. (3, 4, 5). One can see
that they exhibit marked forward peaking, as expected for direct proton
transfer in the $^{7}$Li($^{3}$He,d)$^{8}$Be reaction at the considered high
enough $^{3}$He$^{2+}$ ion bombarding energy of $20$ MeV. Notice that the
first peak of the angular distributions for both three states is described
experimentally for the first time in this work (compare to similar data from
reference \cite{30}). One can also observe in Figs. (3, 4) that the measured
angular distributions for the $2^{+}(16.626)$ and $2^{+}(16.922)$ weakly bound
states have qualitatively similar shapes (although only deuteron forward
angles are concerned here). Furthermore, for this $2^{+}$ doublet of large
isospin mixing the sets of experimental data contained in Figs. (1, 2, 3, 4)
indicate that the lower state at $16.626$ MeV is considerably more strongly
populated than the upper state at $16.922$ MeV. This observation is consistent
with earlier assumptions \cite{9}\ according to which these two $2^{+}$ states
are, respectively, characterized by the $^{7}$Li (g.s.) $+$ p (i.e., $\alpha$
$+$ t $+$ p) and $^{7}$Be (g.s.) $+$ n (or $\alpha$ $+$ $^{3}$He $+$ n)
single-particle model configurations. These properties have been largely
confirmed since then in other previous works both experimentally and via
theoretical (mainly shell-model) calculations. In particular, only the lower
$2^{+}(16.626)$ state was observed in the $^{7}$Li(p,$\gamma$)$^{8}$Be direct
radiative proton capture \cite{9} while the upper $2^{+}(16.922)$ state was
not populated in this reaction. In previous measurements of deuteron spectra
from the $^{7}$Li($^{3}$He,d)$^{8}$Be reaction at \textit{E}$_{lab}$ =
$10.972$ MeV \cite{10} and $15$ MeV \cite{29}, both two $2^{+}$ isospin-mixed
states have been observed to be comparably populated at $\theta_{lab}=$ $40%
{{}^\circ}%
$ while at $10%
{{}^\circ}%
$ and $25%
{{}^\circ}%
$ the lower state at $16.626$ MeV was found to be much more excited than the
$2^{+}(16.922)$ upper state The ratio of the measured differential cross
sections for these two states, $\frac{(d\sigma/d\Omega)_{16.626}}%
{(d\sigma/d\Omega)_{16.922}}$, was found to be an increasing function of the
$^{3}$He ion bombarding energy. Its value derived in the current work is the
highest one ever reported from the $^{7}$Li($^{3}$He,d)$^{8}$Be reaction. This
ratio is higher at small detection angles attaining a maximum value of $58$
for $\theta_{lab}=$ $10%
{{}^\circ}%
$ while a minimum value of $9.7$ is reached at $\theta_{lab}=$ $30%
{{}^\circ}%
$. This behavior was predictable because the proton stripping pattern of the
$2^{+}(16.626$) state enhances its population at small forward angles. Notice
that this ratio is considerably higher in comparison to the peak intensity
ratio, $\frac{I_{16.626}}{I_{16.922}}$, reflecting the relative peak contents
of the two $2^{+}$, $T=0+1$\ isospin-mixed states within the measured deuteron
energy spectra. This is due to the two following facts: \ \ \ \ \ \ \ \ \ \ \ \ \ \ \ \ \ \ \ \ \ \ \ \ \ \ \ \ \ \ \ \ \ \ \ \ \ \ \ \ \ \ \ \ \ \ \ \ \ \ \ \ \ \ \ \ \ \ \ \ \ \ \ \ \ \ \ \ \ \ \ \ \ \ 

(i) the differential cross section ratio corresponds here only to the direct
reaction contribution in the $^{7}$Li($^{3}$He, d)$^{8}$Be reaction; it then
appears that the population of the $2^{+}(16.922)$ upper state of dominant
$^{7}$Be $+$ n single-particle configuration by direct proton transfer is much
less favourable than that of the $2^{+}(16.626)$ lower state at the
investigated $^{3}$He$^{2+}$ ion energy of 20 MeV, while the population of
both two states via the compound nucleus formation mechanism\ is expected to
be drasically\ reduced,\ \ \ \ \ \ \ \ \ \ \ \ \ \ \ \ \ \ \ \ \ \ \ 

ii) the interference between these two $2^{+}$ states, which is indispensable
for a correct treatment of the corresponding deuteron energy spectra, has been
taken into account in the current study; its main effect is to enhance the
ratio of the proton stripping yields to these states as the bombarding energy
increases (see reference \cite{10}).

It was observed, indeed, that at $^{3}$He ion bombarding energies,
E$_{lab}=11$ MeV (see reference \cite{10}), $15$ MeV (see reference
\cite{29}), $20$ MeV (this experiment) and $33$ MeV (see reference \cite{30}),
the $^{7}$Li($^{3}$He,d)$^{8}$Be reaction is increasingly dominated by the
direct proton stripping mechanism essentially populating the $2^{+}$($16.626$)
state, while the $2^{+}$($16.922$) state is much less favourably excited due
to its characteristic $^{7}$Be $+$ n configuration. As a result, the
difference in peak heights for the two $2^{+}$ states is considerably enhanced
but the cross section ratio for the direct reaction component increases much
more with bombarding energy than the peak intensity ratio for the two
states.\textbf{ }Then, the $2^{+}$($16.922$) state has not been obviously
pointed out in\emph{ }the deuteron spectra recorded by Basak et al. \cite{30}
in their study of the $^{7}$Li($\overrightarrow{^{3}\text{He}}$,d)$^{8}$Be
reaction for higher bombarding energy, E$_{lab}$ $=33.3$ MeV; however, a
corresponding shoulder can be seen in the high energy tail of the strong peak
prominent in Fig. 1 (c) of reference \cite{30}, fully attributed by the
authors to the $2^{+}$($16.626$) state. Note also that the compound nucleus
component (probably negligible, indeed, at this high ion bombarding energy)
was not extracted by these authors. To our knowledge, no experimental angular
distribution data for the weaker $2^{+}(16.922$) state of $^{8}$Be has been
reported previously from the $^{7}$Li($^{3}$He,d)$^{8}$Be reaction.\emph{ }As
stated in Section I, only a global experimental angular distribution
exhibiting forward peaking has been reported \cite{31} from this reaction for
the $2^{+}(16.626$) state for \textit{E}$_{lab}=10$ MeV, while the measured
angular distribution for the $2^{+}(16.922$)\ state (not reported in reference
\cite{31}) was claimed to be roughly isotropic, very likely due to a
significant compound nucleus contribution. In previous studies of other direct
reactions involving $^{8}$Be as residual nucleus where the patterns of the
$2^{+}(16.626$) and $2^{+}(16.922$)\ states were experimentally resolved, the
corresponding angular distributions have been generally found to be also
forward peaked and of similar shapes. This trend has been observed in the
$^{10}$B(d,$\alpha$)$^{8}$Be reaction for $7.5$ MeV incident deuterons
\cite{11} where a violation of the isospin selection rule was pointed out with
a total cross section ratio, $\frac{\sigma_{16.626}}{\sigma_{16.922}}=1.15$,
determined without taking into account the interference between these two
$2^{+}$ states, whereas the latter states are expected to be equally populated
in this $T=0$ reaction. Their trends versus bombarding energy was usually
found to be consistent with Marion's single-particle model configuration
assumptions \cite{9}. That is, in reactions where the $^{7}$Li $+$ p
configuration is favoured, the $2^{+}(16.626$) state was observed to be
typically much strongly populated relative to the weaker $2^{+}(16.922$)
state. This was particularly the case in the $^{7}$Li(d,n)$^{8}$Be proton
stripping reaction (see reference \cite{40} and references [1-6] therein). The
measured neutron angular distributions from the latter reaction for high
deuteron bombarding energy \cite{41}, never published in details, were found
to have the following characteristics: (i) in the case of the $2^{+}(16.922$)
state, the neutron angular distribution showed strong forward peaking
consistently with a $\ell=1$ stripping scheme with a differential cross
section value of about $23$ mb/sr at $\theta_{lab}=$ $0%
{{}^\circ}%
$, (ii) that of the $2^{+}(16.922$) upper state was found to be essentially
isotropic with an average cross section value of $\thicksim0.87$ mb/sr, hence
a ratio $\frac{(d\sigma/d\Omega)_{16.626}}{(d\sigma/d\Omega)_{16.922}}%
\simeq26.44$. The two members of the $2^{+}$ isospin-mixed doublet of $^{8}$Be
have also been pointed out in the $^{7}$Li(d, $\alpha\alpha$)n three-body
breakup reaction \cite{15,42}. In the more recent of these studies carried out
for deuteron energies in the range, E$_{lab}=$ $3-6$ MeV \cite{15} where the
above two $2^{+}$ states of the $^{8}$Be were experimentally well resolved,
the measured coincidence spectra indicated strong direct proton population of
the $2^{+}(16.626$) state at neutron emission forward angles consistently with
the assumption \cite{9} of predominant $^{7}$Li(g.s.) $+$ p configuration for
this state. In addition, the assumed interference between the two $2^{+}$,
isospin-mixed states was confirmed experimentally in the latter work
\cite{15}. Besides, in a previous study of the $^{11}$B(p,$\alpha$)$^{8}$Be
reaction at \textit{E}$_{p}=40$ MeV \cite{43} assumed to proceed via the knock
out of an $\alpha-$ particle from $^{11}$B leaving the $^{8}$Be nucleus in the
$^{7}$Li $+$ p (i.e., $\alpha+$ t $+$ p) configuration at this high proton
energy, the angular distributions of both two $2^{+}$, $T=0+1$ isospin-mixed
states were found to be forward peaked and of nearly similar shapes with a
$16.626$/$16.922$ excitation energy ratio of $2.3\pm0.4$. Conversely, the
$2^{+}(16.922$) state was observed to be more strongly populated in reactions
favouring the $^{7}$Be $+$ n single-particle model configuration, such as the
$^{9}$Be(p,d)$^{8}$Be and $^{9}$Be(d,t)$^{8}$Be neutron pick-up reactions for
high particle bombarding energies \cite{44,45,46}.

Finally, the experimental differential cross section for the formation of the
$^{10}$B compound nucleus versus the observation angle proved to be not
important in this $^{7}$Li($^{3}$He,d)$^{8}$Be experiment at $^{3}$He$^{2+}$
ion bombarding energy, E$_{lab}=20$ MeV. This component, likely originating
only from the population of the $2^{+}(16.922$) weaker state, has been pointed
out in the current analysis at only four detection angles. It has been
evaluated within the $^{8}$Be excitation energy region, \textit{E}%
$_{x}=16.626-16.922$ MeV, by inserting into equation \ref{eq.6} the parameter,
$\gamma=N_{c}(\theta)(C^{2}+D^{2})$, for determining the corresponding number
of counts in the deuteron energy spectra. It's observed shape was found to be
roughly isotropic with average values of $\sim0.9$, $0.4$, $0.5$ and $0.8$
mb/sr for $\theta_{lab}=25%
{{}^\circ}%
,30%
{{}^\circ}%
,35%
{{}^\circ}%
$ and $40%
{{}^\circ}%
$, respectively.\emph{ }

\section{Analysis of cross section data, results and discussion}

\subsection{FR-DWBA analysis of angular distribution data}

The measured angular distributions for the two $2^{+}$ bound states at E$_{x}$
= $16.626$ and $16.922$MeV and the $1^{+}$ unbound state at E$_{x}$ = $17.640$
MeV of $^{8}$Be produced in the $^{7}$Li($^{3}$He,d)$^{8}$Be transfer reaction
have been analyzed in terms of the Non Local, Finite Range-DWBA formalism for
direct nuclear reactions \cite{47} with focussing our attention essentially on
the small deuteron forward angles. The calculations have been carried out
using the computer code FRESCO \cite{48} with assuming that the transfer of a
proton occurs from the $^{3}$He$^{2+}$ projectile onto the 1p-shell of the
$^{7}$Li target nucleus. The optical model potential parameters adopted to
describe the distorted waves in the $^{7}$Li $+$ $^{3}$He entrance channel
\cite{49} and the d $+$ $^{8}$Be exit channel \cite{50} are reported in table
II. The bound state wave functions were computed using Saxon-Woods potential
form factors to describe the binding of the proton to the deuteron and $^{7}%
$Li cores in the entrance and exit channels, respectively, while the potential
well depths for the p $+$ d and p $+$ $^{7}$Li systems have been adjusted to
reproduce the corresponding experimental separation energies. For the $1^{+}$
($17.640$) unbound state, the DWBA cross sections were calculated by applying
the procedure described in reference \cite{51}, in which the unbound state
wave function is substituted by that of the $^{7}$Li $+$ p system in a
scattering resonance state. The resonance occurs at an energy for which the
phase shift passes through $\pi/2$ and the wave functions can be calculated by
resolving the radial wave equation at the resonance energy. However, the
oscillatory behavior of the final distorted wave and the unbound state wave
functions induces convergence difficulties in the integral of the reaction
matrix elements. FRESCO deals with unbound states by discretizing the
continuum states in energy bins. A bin wave function is constructed by the
superposition of scattering states within an energy range around the resonance
energy. As the radial partial-wave integral converges very slowly, it must be
extended over several hundreds of femtometers in order to obtain both the
convergence and results independent on the upper value of the cutoff radius
\cite{52}. In the present case, we have used a value of $200$ fm for this parameter.

\begin{center}%
\begin{table}[tbp] \centering
%

\begin{tabular}
[c]{|c|ccccccccccc|}\hline
{\small Elastic scattering} & {\small V}$_{R}$ & {\small r}$_{R}$ &
{\small a}$_{R}$ & {\small W}$_{V}$ & {\small W}$_{D}$ & {\small r}$_{V,D}$ &
{\small a}$_{V,D}$ & {\small V}$_{s.o.}$ & {\small r}$_{s.o.}$ &
{\small a}$_{s.o.}$ & {\small r}$_{c}$\\
& {\small (MeV)} & {\small (fm)} & {\small (fm)} & {\small (MeV)} &
{\small (MeV)} & {\small (fm)} & {\small (fm)} & {\small (MeV)} &
{\small (fm)} & {\small (fm)} & {\small (fm)}\\\hline
$^{{\small 3}}${\small He + }$^{{\small 7}}${\small Li \ \cite{49} } &
{\small 146.9} & {\small 1.39} & {\small 0.684} & {\small 29.1} & {\small 0} &
{\small 1.912} & {\small 0.407} & {\small 5.21} & {\small 1.426} &
{\small 0.211} & {\small 1.4}\\
{\small d + }$^{8}${\small Be \cite{50}} & {\small 90} & {\small 0.9} &
{\small 0.9} & {\small 10} & {\small 6.25} & {\small 1.6} & {\small 0.8} &
{\small 5} & {\small 1.6} & {\small 0.8} & {\small 1.3}\\\hline
\end{tabular}
%

\end{table}%

\end{center}

\subsection{Results and discussion}

\subsubsection{Angular distributions}

The DWBA-generated theoretical curves corresponding to best fits to the $^{7}%
$Li($^{3}$He,d)$^{8}$Be angular distribution experimental data for the three
studied states of $^{8}$Be (the $2^{+}(16.626$) and $2^{+}(16.922$), $T=0+1$
isospin-mixed states and the $1^{+}(17.64$), $T=1$ state) are also plotted in
Figs. (3, 4, 5). One observes that the cross section experimental data for the
strongly populated $2^{+}$ state at $16.626$ MeV are satisfactorily reproduced
by the DWBA-calculated curve, which further confirms the validity of the
assumption \cite{9} that the structure of this state is dominated by the
$^{\emph{7}}$Li $+$ p single-particle model configuration. Those for the
weaker $2^{+}$ ($16.922$) state also appear to be well accounted for by our
DWBA calculation.$\ $The good description by this theory of the angular
distribution data from the $^{7}$Li($^{3}$He,d)$^{8}$Be proton stripping
reaction for this state means that the structure of the latter also involves
the $^{\emph{7}}$Li $+$ p single-particle model configuration and that it
cannot be of pure $^{7}$Be $+$ n configuration as assumed earlier in reference
\cite{9}. Therefore, by fitting the corresponding lowest forward angle cross
section data, one can deduce a reliable\ proton spectroscopic information for
this state from the $^{7}$Li($^{3}$He,d)$^{8}$Be proton stripping reaction at
the relatively high $^{3}$He$^{2+}$ ion bombarding energy of $20$ MeV.

Besides, the angular distribution experimental data for the $1^{+}(17.640)$
unbound state are less well reproduced by our DWBA calculation, as can be seen
in Fig. 5. It must be noted that several theoretical fits to the angular
distribution experimental data for this state have been tried without
simultaneously accounting for all the measured cross section values. Finally,
the following procedure was adopted : first, a theoretical best fit was
obtained by considering all forward angle experimental data points; then, a
second best fit was derived with ignoring the smallest angle data point that
did not follow the calculated curve. Two corresponding $C^{2}S$ values were
thus determined and their mean value was adopted. Consequently, a large
uncertainty (standard deviation) of $\sim43$ $\%$ has been assigned to the
derived proton spectroscopic factors for this state.

\subsubsection{Proton spectroscopic factors}

The following expression relating the experimental differential cross section
to the DWBA theoretical counterpart was used in order to deduce the proton
spectroscopic factors for the three studied states of $^{8}$Be%

\begin{equation}
\left(  \frac{d\sigma}{d\Omega}\right)  _{\exp}=4.43C^{2}\frac{2J_{f}%
+1}{2J_{i}+1}\underset{n\ell j}{\sum}\frac{S_{n\ell j}}{2j+1}\left(
\frac{d\sigma_{n\ell j}(\theta)}{d\Omega}\right)  _{DWBA}. \label{eq.7}%
\end{equation}
In this formula, the factor of $4.43$ is the commonly admitted value \cite{53}
to describe the $\left\langle d\otimes p\right\vert $ $^{3}He\rangle$ overlap
function in the ($^{3}$He, d) reaction, the $C$ factor is the Clebsch-Gordon
coefficient coupling the isospins of the target and final nuclei and the
transferred particle (in the present case of the $^{7}$Li($^{3}$He,d)$^{8}$Be
reaction, $C^{2}=1/2$ for the final states, $T=0,1$), ($J_{i}$, $J_{f}$) are
the respective spins of the target nucleus and the studied state in the
residual nucleus, and $S_{n\ell j}$ is the spectroscopic factor for the
transfer of a proton onto a shell-model orbit of the $^{7}$Li target
characterized by the $\{n\ell j\}$ set of quantum numbers (i.e., the principal
quantum number, $n$, the orbital and total angular momentum quantum numbers,
$\ell$ and $j$) to form the final state of the $^{8}$Be nucleus. Due to their
spin and parity characteristics, $J_{f}^{\pi}=2^{+}$ and $1^{+}$, the final
states of $^{8}$Be considered in this study can be populated via both
$1p_{1/2}$ and $1p_{3/2}$ captures. Then, for a dominating $1p_{3/2}$
transition, equation \ref{eq.7} rewrites as%

\begin{equation}
\left(  \frac{d\sigma}{d\Omega}\right)  _{\exp}=4.43C^{2}\frac{2J_{f}%
+1}{4(2J_{i}+1)}\left\{  1+2\left(  \frac{S_{1p_{1/2}}}{S_{1p_{3/2}}}\right)
\frac{\left(  d\sigma/d\Omega\right)  _{1p_{1/2}}^{DWBA}}{\left(
d\sigma/d\Omega\right)  _{1p_{3/2}}^{DWBA}}\right\}  S_{1p_{3/2}}\left(
\frac{d\sigma}{d\Omega}\right)  _{1p_{3/2}}^{DWBA}, \label{eq. 8}%
\end{equation}

where $S_{global}$ is defined as a global spectroscopic factor by%

\begin{equation}
S_{global}=\left\{  1+2\left(  \frac{S_{1p_{1/2}}}{S_{1p_{3/2}}}\right)
\frac{\left(  d\sigma/d\Omega\right)  _{1p_{1/2}}^{DWBA}}{\left(
d\sigma/d\Omega\right)  _{1p_{3/2}}^{DWBA}}\right\}  S_{1p_{3/2}}. \label{9}%
\end{equation}

Since the shapes of the $1p_{1/2}$ and $1p_{3/2}$ capture cross sections
calculated with the FRESCO program are quasi identical, their ratios in the
latter equation have been considered as constants over the whole studied
angular range, $\theta_{lab}=0%
{{}^\circ}%
-50%
{{}^\circ}%
$. The values of these ratios are reported in table III together with the
$S_{1p_{1/2}}/S_{1p_{3/2}}$ partial spectroscopic factor ratios derived here
for the three studied states of $^{8}$Be based on Cohen and Kurath's shell
model calculations \cite{54} (see below). Indeed, neither the $S_{1p_{1/2}}$
and $S_{1p_{3/2}}$ partial spectroscopic factors nor their ratio,
$S_{1p_{1/2}}/S_{1p_{3/2}}$, can be directly derived from the DWBA analysis.
Fortunately, Cohen and Kurath have performed shell model calculations
\cite{54} where spectroscopic factors have been calculated for single nucleon
stripping on the $1$p shell of a target nucleus with mass number, $A=7$,
leading to the population of states with properties, $J_{f}^{\pi}=2^{+}$
$(T=0,1)$ and $1^{+}$ $(T=1)$, which we have identified to the $2^{+}%
$\ doublet at $E_{x}$ $=16.626$ MeV and $16.922$ MeV and the $1^{+}$\ state at
$E_{x}$ $=17.640$ MeV in $^{8}$Be, respectively. We have therefore used Cohen
and Kurath's results to derive the $S_{1p_{1/2}}/S_{1p_{3/2}}$ ratios reported
in table III, that we have considered in equation $9$ for deducing
experimental values of the $S_{1p_{1/2}}$ and $S_{1p_{3/2}}$ partial
spectroscopic factors for the $1p$ shell. Cohen and Kurath \cite{54} have also
calculated $S_{1p_{1/2}}$ and $S_{1p_{3/2}}$ spectroscopic factor values
separately for each isospin state. Then, the $(S_{1p_{1/2}}/S_{1p_{3/2}%
})_{C.K.}$ ratio following these authors \cite{54} for the pure $T=1$ isospin
state at $17.640$ MeV was directly deduced, while the same ratio for the
$T=0+1$ isospin-mixed states at $16.626$ MeV and $16.922$ MeV were derived
using the spectroscopic factor values calculated as follows for $j=1/2$ and
$3/2$:%

\begin{align}
S_{1p_{j}\text{ }C.K.}(16.626\text{ }MeV)  &  =AS_{1p_{j}\text{ }%
C.K.}(T=0)+BS_{1p_{j}\text{ }C.K.}(T=1),\label{eq.  10}\\
S_{1p_{j}\text{ }C.K.}(16.922\text{ }MeV)  &  =\left\vert BS_{1p_{j}\text{
}C.K.}(T=0)-AS_{1p_{j}\text{ }C.K.}(T=1)\right\vert ,\nonumber
\end{align}
where $A=0.797$ and $B=0.604$ as defined in Section II.

\begin{center}%
\begin{table}[tbp] \centering
%

\begin{tabular}
[c]{|c|c|c|c|}\hline
${\small J}^{\pi}({\small E}_{x},{\small MeV)}$ & {\small T} & $\left(
d\sigma/d\Omega\right)  _{1p_{1/2}}^{DWBA}{\small /}\left(  d\sigma
/d\Omega\right)  _{1p_{3/2}}^{DWBA}$ & ${\small (S}_{1p_{1/2}}{\small /S}%
_{1p_{3/2}}{\small )}_{C.K.}$\\
{\small \ \ \ \ \ \ \ \ \ \ \ \ \ \ \ \ \ \ } & {\small \ \ \ \ \ \ } &
{\small This work} & {\small Cohen and Kurath \cite{54}\ }\\\hline
${\small 2}^{+}{\small (16.626)}$ & ${\small 0+1}$ & ${\small 0.660\pm0.034}$
& ${\small 0.2262}$\\
${\small 2}^{+}{\small (16.922)}$ & ${\small 0+1}$ & ${\small 0.680\pm0.043}$
& ${\small 0.4190}$\\
${\small 1}^{+}{\small (17.640)}$ & ${\small 1}$ & ${\small 1.036\pm0.019}$ &
${\small 0.3857}$\\\hline
\end{tabular}
%

\end{table}%

\end{center}

The derived values of the absolute proton spectroscopic factor, $C^{2}%
S_{global}$, for the three studied states of $^{8}$Be are reported in table IV
where they are compared to the only experimental values available in the
literature deduced by Basak et al. \cite{30} from the $^{7}$%
Li($\overrightarrow{^{3}He}$, d)$^{8}$Be transfer reaction for a bombarding
energy, \textit{E}$_{lab}=33.3$ MeV. We recall here that probably due to
insufficient\emph{ }experimental energy resolution, the peaks for the
$2^{+\text{ }}$($16.626$) and $2^{+\text{ }}$($16.922$) states in the deuteron
energy spectra from that experiment \cite{30} were not separated. Furthermore,
considering the latter state to be of pure $^{7}$Be $+$ n single-particle
model configuration, these authors have neglected its weak excitation in the
$^{7}$Li($\overrightarrow{^{3}He}$, d)$^{8}$Be reaction and, then, have likely
over-estimated the population of the stronger $2^{+\text{ }}$($16.626$) state
in this reaction, which could explain why their extracted $C^{2}S_{global}$
value is a factor $1.61$ times our result.

In order to compare our experimental $C^{2}S_{global}$ values to corresponding
theoretical counterparts, we have used equation $9$ with the spectroscopic
factor values calculated by Cohen and Kurath \cite{54}. These authors obtained
the following global spectroscopic factors for each isospin $T-$ value:
$0.289$ $(J^{\pi}=1^{+},T=1),$ $0.575$ $(J^{\pi}=2^{+},T=0)$ and $0.525$
$(J^{\pi}=2^{+},T=$ $1)$. Then, the Cohen and Kurath $C^{2}S_{global}$ values
reported in table IV for the $T=0+1$ isospin - mixed states have been
calculated using the relations%

\begin{align}
C^{2}S_{global\text{ }C.K.}(16.626\text{ }MeV)  &  =AC^{2}S_{global\text{
}C.K.}(T=0)+BC^{2}S_{global\text{ }C.K.}(T=1),\label{eq.  11}\\
C^{2}S_{global\text{ }C.K.}(16.922\text{ }MeV)  &  =\left\vert BC^{2}%
S_{global\text{ }C.K.}(T=0)-AC^{2}S_{global\text{ }C.K.}(T=1)\right\vert
.\nonumber
\end{align}

Besides, the results derived here for the proton ($S_{1p_{1/2}}$,
$S_{1p_{3/2}}$) partial spectroscopic factors of the three studied states of
$^{8}$Be are reported in table V where they are compared to Cohen and Kurath's
shell model predictions \cite{54}.

\begin{center}%
\begin{table}[tbp] \centering
%

\begin{tabular}
[c]{|c|cccc|}\hline
${\small J}^{\pi}${\small (E}$_{x}${\small ,MeV)} &
\multicolumn{4}{|c|}{${\small C}^{2}{\small S}_{global}${\small \ results}}\\
{\small \ \ \ \ \ \ \ \ \ \ \ \ \ } & {\small This work\ } & {\small Basak
\cite{30}} & {\small Cohen and Kurath \cite{54}} & F. C. Barker
{\small \cite{13}}\\\hline
${\small 2}^{+}{\small (16.626)}$ & ${\small 0.650\pm0.097}$ &
${\small 1.05\pm0.10}$ & ${\small 0.775}$ & ${\small 0.475}$\\
${\small 2}^{+}{\small (16.922)}$ & ${\small 0.0195\pm0.007}$ &
\footnote{{\small Unseparated from the 16.626 MeV state.}} & ${\small 0.071}$
& ${\small 0.005}$\\
${\small 1}^{+}{\small (17.640)}$ & ${\small 0.439\pm0.190}$ &
${\small 0.30\pm0.15}$ & ${\small 0.289}$ & ${\small 0.145}$\footnote{From
resonance reaction}\\\hline
\end{tabular}
%

\end{table}%

\end{center}

One can observe, first, in both two tables (IV, V) that our spectroscopic
factors for the weakly excited $2^{+}(16.922$) state are remarkably smaller
than those for the strongly populated $2^{+}(16.626$) state in the $^{7}%
$Li($^{3}$He,d)$^{8}$Be reaction, in overall consistency with shell model
predictions \cite{13,54}. These results, which can be expected from the
corresponding patterns of these two $2^{+}$states in the recorded deuteron
energy spectra (see Figs. (1, 2)) and from the measured angular distributions
(see Figs. (3, 4)) obviously confirm Marion's assumptions \cite{9, 10} that
the $^{7}$Be $+$ n and $^{\emph{7}}$Li $+$ p single-particle model
configurations dominate, respectively, the structures of these two $2^{+},$
$T=0+1$ isospin-mixed\ states. As can be seen in table IV, our new
$C^{2}S_{global}$ experimental value for the strongly excited $2^{+}$
$(16.626)$ state is in good agreement with Cohen and Kurath's \cite{54} shell
model-predicted one. It is also very consistent with the value derived by
Barker \cite{13} who has performed shell model calculations for $^{8}$Be
states using single-particle wave functions in a harmonic oscillator
potential. Concerning the $S_{1p_{1/2}}$ and $S_{1p_{3/2}}$ partial
spectroscopic factors (see table V), a similar agreement is also observed
between our values for the $2^{+}$ $(16.626)$ state and corresponding Cohen
and Kurath's shell-model-predicted ones \cite{54}. In the case of the
$2^{+}(16.922)$ weaker state, however, our results are substantially lower
than the predictions of Cohen and Kurath: by a factor of\ $\sim3.64$
concerning $C^{2}S_{global}$ (see table IV) and\ by more than one order of
magnitude concerning the $S_{1p_{1/2}}$ and $S_{1p_{3/2}}$ partial
specroscopic factors (see table V). Note that, in contrast, our experimental
value of $C^{2}S_{global}$ for this state (see table IV) is a factor $3.9$
higher than that derived by Barker via shell model calculation \cite{13}.

It is thus clearly pointed out here that the population of the $2^{+}(16.922$)
state in the $^{7}$Be $+$ n single-particle model configuration is minor in
the $^{7}$Li($^{3}$He,d)$^{8}$Be proton stripping reaction. Cohen and Kurath's
effective-interaction calculations \cite{54} are charge independent. Then, the
corresponding shell model spectroscopic factors for the $2^{+}(16.922$) state
should be more appropriately compared to experimental counterparts resulting
from a DWBA analysis of angular distribution data for neutron transfer
reactions populating this state in its dominant $^{7}$Be $+$ n configuration,
like the $^{7}$Be(d,p)$^{8}$Be and $^{7}$Be(t,d)$^{8}$Be stripping reactions
or the $^{9}$Be(p,d)$^{8}$Be and $^{9}$Be(d,t)$^{8}$Be pick-up reactions.
Indeed, absolute neutron $C^{2}S\ $values derived for the latter state from
the $^{7}$Be(d,p)$^{8}$Be and $^{9}$Be(d,t)$^{8}$Be reactions can be found in
references \cite{45,46}, respectively, where they are compared to experimental
values from earlier works and to Cohen and Kurath's shell-model predictions
\cite{54}.

\begin{center}%
\begin{table}[tbp] \centering
%

\begin{tabular}
[c]{|c|cc|cc|}\hline
${\small J}^{\pi}{\small (E}_{x}{\small ,MeV)}$ &
\multicolumn{2}{|c|}{{\small This work\ }} &
\multicolumn{2}{|c|}{{\small Cohen and Kurath \cite{54}\ }}\\
{\small \ \ \ \ \ \ \ \ \ \ \ \ \ } & ${\small S}_{1p_{1/2}}$ & ${\small S}%
_{1p_{3/2}}$ & $\ \ \ \ {\small S}_{1p_{1/2}}$ \ \  & ${\small S}_{1p_{3/2}}%
$\\\hline
${\small 2}^{+}{\small (16.626)}$ & ${\small 0.226\pm0.041}$ &
${\small 1.001\pm0.184}$ & ${\small 0.270}$ & ${\small 1.194}$\\
${\small 2}^{+}{\small (16.922)}$ & ${\small 0.010\pm0.001}$ &
${\small 0.025\pm0.003}$ & ${\small 0.134}$ & ${\small 0.321}$\\
${\small 1}^{+}{\small (17.640)}$ & ${\small 0.188\pm0.085}$ &
${\small 0.488\pm0.222}$ & ${\small 0.124}$ & ${\small 0.321}$\\\hline
\end{tabular}
%

\end{table}%

\end{center}

Concerning the $1^{+}(17.640)$, $T=1$\ state of $^{8}$Be, one can observe in
tables (IV, V) that the proton $C^{2}S_{global}$ and ($S_{1p_{1/2}%
},S_{1p_{3/2}}$) values derived in this work are very consistent both with the
experimental value of Basak et al. \cite{30} and with Cohen and Kurath's
\cite{54} shell model predictions with all results being comprised within the
experimental error bares. One must note, in passing, the observed very good
agreement between the $C^{2}S_{global}$ value derived for this state by Basak
et al. \cite{30} and the shell model-predicted value of Cohen and Kurath
\cite{54}. The present DWBA proton spectroscopic factor results for the
$2^{+}$, $T=0+1$ isospin-mixed doublet in $^{8}$Be appear to be essentially
new since while no previous $C^{2}S_{global}$ experimental value from the
$^{7}$Li($^{3}$He,d)$^{8}$Be transfer reaction has been reported in the
literature for the weaker $2^{+}$($16.922$) state, the unique value deived by
Basak et al. \cite{30} for the $2^{+}$($16.626$) strongly excited state is not
well consistent with shell model predictions \cite{13,54}. Note, in addition,
that no experimental data from this reaction do exist in the previous
literature concerning the proton $S_{1p_{1/2}}$ and $S_{1p_{3/2}}$\ partial
spectroscopic factors for both three states of $^{8}$Be considered in this study.

\subsubsection{$\gamma_{p}^{2}$ $(a)$ proton reduced widths}

With the knowledge of the proton spectroscopic factors inferred in our DWBA
analysis, the proton reduced widths versus the $p+^{7}Li$ system channel
radius can be derived from basic nuclear physics relations. For this purpose,
the used expression was that defined in \cite{55,56}, summed over the
$1p_{1/2}$ and $1p_{3/2}$ proton capture shells of $^{8}$Be in terms of the
individual spectroscopic factors, i.e.,
\begin{equation}
\gamma_{p}^{2}\text{ (}a\text{)}=\frac{\hbar^{2}}{2\mu\text{ }a}\left[
S_{1p_{1/2}}\left\vert u_{1p_{1/2}}(a)\right\vert ^{2}+S_{1p_{3/2}}\left\vert
u_{1p_{3/2}}(a)\right\vert ^{2}\right]  , \label{12}%
\end{equation}
where $\mu$ is the reduced mass of the $p+^{7}Li$ system and $u_{n\ell j}(a)$
is the relative motion radial wave function calculated at channel radius%

\begin{equation}
a=a_{0\text{ }}(A_{p}^{1/3}+A_{^{7}Li}^{1/3}).\label{13}%
\end{equation}
The obtained $\gamma_{p}^{2}$ $(a)$ values for the three studied states of
$^{8}$Be at $16.626$, $16.922$ and $17.640$ MeV excitation energies are
reported in table VI where they are compared to counterparts reported by
Barker from an R-Matrix analysis \cite{34} of improved (ancient) cross section
experimental data for the $^{7}$Li(p,$\alpha$)$^{4}$He reaction (for the
$2^{+}$($16.626$) state) or following shell model calculations (for the
$2^{+}$($16.922$) state \cite{34} and the $1^{+}(17.640)$ state \cite{14}).
Notice, first, that for each of the three studied states of $^{8}$Be, notably
the $1^{+}(17.640)$ state, our derived DWBA\ proton reduced widths versus the
channel radius, $\gamma_{p}^{2}$ $(a)$, approximately remain in the same order
of magnitude and that they are consistent with Barker's shell model or
R-Matrix analysis derived values \cite{14,34} (see table VI). As can be seen
in this table, the $\gamma_{p}^{2}$ $(a)$ values derived here for the strongly
populated $2^{+}$($16.626$) state in the $^{7}$Li($^{3}$He,d)$^{8}$Be proton
stripping reaction are higher by more than one order of magnitude (precisely
by factors of $35.25$, $34.48$ and $34.17$, respectively for $a=$ $3.64$,
$4.22$ and $5.00$ fm) than those obtained for the weaker $2^{+}$($16.922$)
excited state. Again, this observation reflects the fact that the structures
of these two $2^{+}$, $T=0+1$ isospin-mixed states of $^{8}$Be are indeed
respectively dominated by the $^{7}$Li $+$ p and $^{7}$Be $+$ n
single-particle model configurations \cite{9, 10}. Comparatively, the
corresponding $\gamma_{p}^{2}$ $(a)$ values reported by Barker \cite{34} from
R-Matrix analysis (for the $2^{+}$($16.626$) state) or shell model calculation
(for the $2^{+}$($16.922$) state) are within a ratio of $60$. Performing later
an R-Matrix analysis of more recent experimental data for the $^{7}%
$Li(p,$\alpha$)$^{4}$He reaction, Barker \cite{57} has obtained a $\gamma
_{p}^{2}$ value for the $2^{+}$($16.626$) state in agreement with our
corresponding DWBA result but his derived $\gamma_{p}^{2}$ value for the
$2^{+}$($16.922$) state substantially differs from our DWBA counterpart (see
column $7$ of table VI for channel radius, $a=4.22$ fm). The\ $\gamma_{p}$
($a$) proton reduced width amplitudes or the related $\Gamma_{p}$ proton
partial widths are commonly used as free fit parameters in this type of
theoretical analysis \cite{34,57}. In the case of unbound states, the proton
partial width expresses as $\qquad\qquad\qquad\qquad\qquad$%
\begin{equation}
\Gamma_{p}=2P_{\ell}(E)\times\gamma_{p}^{2},\label{14}%
\end{equation}
where $P_{\ell}(E)$ is the Coulomb barrier penetration factor for the involved
$\ell-$ partial wave. However, for sub-threshold states (here the $2^{+\text{
}}$doublet of $^{8}$Be), $P_{\ell}(E)$ cannot be calculated at the
corresponding negative resonance energies, and $\Gamma_{p}$ can be only
estimated via several approximate methods, usually in terms of the proton
spectroscopic factor times a single-particle model proton width (see \cite{58}
and references therein). The derived values of $\gamma_{p}^{2}$ (then of
$\gamma_{p}$ and $\Gamma_{p}$) generally depend on the adopted reaction
channel radius and can be also sensitive to the experimental data sets
considered in the R-Matrix analysis \cite{34,57} for fixed value of $a$.
Besides, one can observe in table VI that our DWBA-derived values of
$\gamma_{p}^{2}$ $(a)$ for the $2^{+}$($16.626$) strongly excited state in the
$^{7}$Li($^{3}$He,d)$^{8}$Be proton stripping reaction show to be higher than
those for the $1^{+}$($17.640$) state by factors of only $1.49$, $1.62$ and
$3.49$ for $a=$ $3.64$, $4.22$ and $5.00$ fm, respectively, which is an
indication that the structure of the latter narrow $T=1$ state appears to be
substantially featured by the $^{7}$Li $+$ p single-particle model
configuration, consistently with Marion's earlier assumption \cite{40}.
Finally, our $\gamma_{p}^{2}$ $(a)$ results for the three studied states of
$^{8}$Be are indeed very concordant with previous observations, as reported in
reference \cite{40}.

As stated in Section I, our DWBA proton reduced widths could be very useful
for quantifying the contributions of the studied $^{8}$Be states in various
nuclear physics or nuclear astrophysics topics via interaction processes
involving this nucleus. In this respect, a careful, new and thorough R-Matrix
analysis of up-dated, appropriately normalised angular distribution and
integrated cross section experimental data sets available in the literature
both via direct \cite{33,59,60} and indirect \cite{61,62,63} measurement
methods for the $^{7}$Li(p,$\alpha$)$^{4}$He hydrogen burning reaction of main
astrophysical concern with considering the current DWBA results appears to us
as highly desirable.

\ \ \ \ \ \ \ \ \ \ \ \ \ \ \ \ \ \ \ \ \ \ \ \ \ \ \ \ \ \ \ \ \ \ 

\begin{center}
\ \ \ \ \ \ \ \ \ \ \ \ \ \ \ \ \ \ \ \ \ \ \ \ \ \ \ \ \ \ \ \ \ \ \ \ \
\begin{table}[tbp] \centering
%

\begin{tabular}
[c]{c|c|c|c|ccc|}\cline{2-7}
& \multicolumn{6}{|c|}{${\small \gamma}_{p}^{2}{\small \ (MeV)}$}\\\cline{2-7}
& \multicolumn{3}{|c|}{{\small Present work}} & \ {\small Ref.\cite{14} \ } &
\ {\small Ref. \cite{34} \ } & \ {\small Ref. \cite{57} \ }\\\cline{2-7}%
\cline{5-7}
& ${\small a=3.64}$ ${\small fm}$ & ${\small a=4.22}$ ${\small fm}$ &
${\small a=5.00}$ ${\small fm}$ & \multicolumn{3}{|c|}{${\small a=4.22}$
${\small fm}$}\\\hline
\multicolumn{1}{|c|}{${\small 2}^{+}{\small (16.626)}$} & ${\small 2.08}$ &
${\small 1.54}$ & ${\small 0.991}$ &  & ${\small 0.896}$ & ${\small 1.659}$\\
\multicolumn{1}{|c|}{${\small 2}^{+}{\small (16.922)}$} & ${\small 0.059}$ &
${\small 0.046}$ & ${\small 0.029}$ &  & ${\small 0.015}$ & ${\small 0.213}$\\
\multicolumn{1}{|c|}{${\small 1}^{+}{\small (17.640)}$} & ${\small 1.4}$ &
${\small 0.95}$ & ${\small 0.284}$ & ${\small 0.321}$ &  & \\\hline
\end{tabular}

\ \ \ \ \ \ \ \ \ \ \ \ \ \ \ \ \ \ \ \ \ \ \ \ \ \ \ \ \ \ \ \ \ \ \ \ \
\end{table}%

\end{center}

\section{Summary and conclusions}

In this work, a complete and up-to-date status of the spectroscopic properties
of the $^{8}$Be $2^{+}$ ($16.626$) and $2^{+}$ ($16.922$), $T=0+1$
isospin-mixed state of crucial interest in nuclear astrophysics has been
addressed for several purposes, mainly regarding the nucleosynthesis of $^{7}%
$Li \cite{17}. In this respect, we have measured the corresponding angular
distributions in the $^{7}$Li($^{3}$He,d)$^{8}$Be transfer reaction for $^{3}%
$He$^{2+}$ ion bombarding energy, \textit{E}$_{lab}=$ $20$ MeV,\ covering the
forward angular range, $5%
{{}^\circ}%
\leq\theta_{lab}\leq50%
{{}^\circ}%
$, in $5%
{{}^\circ}%
$ steps. A clear separation in the recorded deuteron energy spectra of the
peaks associated with these two $2^{+}$\ states has been achieved, thanks to
the high energy resolution of the detection system used in the experiment. As
expected, the $2^{+}$ ($16.626$)\ state, essentially featured by the $^{7}$Li
$+$ p single-particle model configuration \cite{9,10}, showed to be much more
strongly populated than the $2^{+}$ ($16.922$)\ state confirmed to be
primarily of the $^{7}$Be $+$ n configuration, pointed out here to be minor in
this reaction for the considered kinematics. In addition, the angular
distribution for the narrow $1^{+}(17.640)$, $T=1$ state was measured.\ The
experimental data for the above three states of $^{8}$Be have been analysed in
the framework of the non-local, FR-DWBA theory by considering proton captures
from the $^{3}$He$^{2+}$ projectiles onto the $1$p$_{3/2}$ and$\ 1$p$_{1/2}$
shells of the final nuclei. Updated values of the proton absolute and
partial\ spectroscopic factors as well as of the\ proton reduced widths have
been extracted for these three states. Within the DWBA uncertainties, our
$C^{2}S_{global}$ value for the $2^{+}$ ($16.626$) strongly excited state was
found to be in good agreement with shell-model predictions \cite{13,54}, while
the unique previous experimental value for this state (see reference
\cite{30}) was found to exceed our result by $\sim40\%$ and to lie
substantially above theoretical predictions. In contrast, our $C^{2}%
S_{global}$ result for the $2^{+}(16.922)$ weakly populated state was found to
be lower by a factor of \ $\sim3.64$ than Cohen and Kurath's \cite{54} shell
model value and in excess by a factor of $3.9$ relative to Barker's \cite{13}
theoretical value. Similar observations evenly hold concerning the derived
($S_{1p_{1/2}}$, $S_{1p_{3/2}}$) partial spectroscopic factors for the three
studied states, i.e., fair agreement between our DWBA values for the $2^{+}%
$($16.626$) and $1^{+}$($17.640$) states and their corresponding shell model
counterparts \cite{54}, and large differences concerning the $2^{+}$($16.922$)
state of predominant $^{7}$Be $+$ n single-particle model configuration.
Besides, the $\gamma_{p}^{2}$ $(a)$ proton reduced\ widths versus the $^{7}$Li
$+$ p channel radius have been deduced from our DWBA analysis of the
experimental angular distribution data. The derived values for a $^{7}$Li $+$
p channel radius, $a=4.22$ fm, were found to be consistent both with earlier
shell model calculations for the $2^{+}$($16.922$) state \cite{34} and the
$1^{+}$($17.640$) state \cite{14}, and in good agreement for the $2^{+}%
$($16.626$) state with a counterpart value derived by Barker \cite{57} from an
R-Matrix analysis of more recent experimental data for the $^{7}$Li(p,$\alpha
$)$^{4}$He resonant reaction of astrophysical concern. Then, the\textbf{
}$\gamma_{p}$\textbf{ }$(a)$\textbf{ }reduced width amplitudes (or
the\ related $\Gamma_{p}(a)$ proton partial widths) can be used in the
R-Matrix analysis of cross section experimental data for this reaction. The
present DWBA\ results concerning the spectroscopic parameters of the$\ $three
studied excited states of $^{8}$Be could be further confirmed via a similar
study of the $^{7}$Li(d,n)$^{8}$Be proton stripping reaction for which cross
section angular distribution data for high deuteron bombarding energy are very
lacking in the literature. In addition, the measurement of the angular
distributions for the $^{7}$Li($^{3}$He,d)$^{8}$Be transfer reaction at $^{3}%
$He ion bombarding energy, E$_{lab}=$ $33$ MeV, could permit to compare in the
same kinematics conditions the obtained results to those of Basak et al.
\cite{30}, and also to check if the ($A$, $B$) amplitudes of the direct
reaction component in equation \ref{eq.4} are completely independent on the
reaction bombarding energy. The DWBA results inferred here can be pertinently
used to address several open questions in nuclear physics and nuclear
astrophysics concerning interaction processes involving the $^{7}$Li and/or
$^{8}$Be isotopes, as emphasised in Section I. In particular, a new R-Matrix
analysis of relevant, well selected and appropriately normalised cross section
experimental data for the $^{7}$Li(p,$\alpha$)$^{4}$He hydrogen burning
reaction with considering the present $\gamma_{p}$ $(a)$\ DWBA-extracted
results appeared to us as highly necessary, and has been recently undertaken
by members \cite{64} of our group. Thanks to the R-Matrix analyses of the
available large corpus of precise experimental data sets accumulated during
the last four decades for the $^{7}$Li(p,$\alpha$)$^{4}$He reaction (involved
in BBN and stellar nucleosyntheses) both via direct and indirect measurement
methods, the associated astrophysical $S(E)-$factor and $N_{A}<\sigma v>$ rate
can be determined with very high accuracy at thermonuclear energies. The
corresponding results can be very helpful to obtain reliable informations on
the abundance of lithium and beryllium isotopes in different astrophysical
sites. On the other hand, these results strongly support existing clear
experimental evidences for ($^{8}$Be) compound nucleus formation in the $^{7}%
$Li(p,$\alpha$)$^{4}$He fusion reaction at thermonuclear energies, in contrast
with previous indications \cite{32} according to which a three-nucleon
transfer reaction mechanism would dominate the proton sub-Coulomb energy
regime of this reaction.

\textbf{ACKNOWLEDGMENTS :}

One of us (S. O.) is indebted to Dr A. Coc from the CSNSM/Orsay for helpful
discussion, and to Prof M. Debiane from the USTHB/Algiers for providing some
bibliography items.

\textbf{Tables captions: }

TABLE I: Best fit values of parameters entering in equation \ref{eq.4} fitted
to the experimental deuteron energy spectra from this work compared to those
obtained by Marion et al. \cite{10}.\textbf{\ }\ \ \ \ 

TABLE II: Optical model potential parameters for the $^{3}$He + $^{7}$Li

\cite{49} and d + $^{8}$Be \cite{50} elastic scattering reaction channels.

TABLE III: Ratios of the 1p$_{1/2}$ to 1p$_{3/2}$ shells partial DWBA
differential cross sections and Cohen and Kurath shell model-calculated
\cite{54} spectroscopic factors for the three studied states of $^{8}$Be.

TABLE IV: Experimental $C^{2}S$ global spectroscopic factor values for the
three studied states of $^{8}$Be inferred in this work from our DWBA analysis
of $^{7}$Li($^{3}$He,d)$^{8}$Be experimental data compared to Basak et al.
experimental values \cite{30} and to shell model calculated \cite{13,54} counterparts.

TABLE V: Experimental and shell model-predicted \cite{54} partial
spectroscopic factor values for the three studied states of $^{8}$Be.

TABLE VI: $\gamma_{p}^{2}$ $(a)$ proton-reduced widths versus the p $+^{7}$Li
channel radius deduced in this work from our DWBA analysis of $^{7}$Li($^{3}%
$He,d)$^{8}$Be experimental data compared to shell model-predicted
\cite{14,34} and R-matrix \cite{57} counterparts.\bigskip

\textbf{Figures captions:}

FIG. 1: Part of the deuteron energy (position) spectrum recorded at
$\theta_{\text{lab}}=5%
{{}^\circ}%
$ in our $^{7}$Li($^{3}$He,d)$^{8}$Be experiment.

FIG. 2 : Portion of the deuteron energy spectrum\ for $\theta_{\text{lab}}=35%
{{}^\circ}%
$ showing the $2^{+}$ (16.626) and $2^{+}$\ (16.922) states of $^{8}$Be very
well resolved due to the high energy resolution of the detection system used
in this experiment. The solid curve represents the best fit to the
experimental data points obtained from the Breit-Wigner two-level expression
\ref{eq.4}. The dotted curves represent the separate components associated
with these two $2^{+}$ states\ while the dashed curve describes their
interference (last term in \ref{eq.4}).

FIG. 3 : Experimental cross section angular distributions (scatter points) of
the $2^{+}$ ($16.626$) state of $^{8}$Be.\ The solid curve represents the best
theoretical fit to experimental data from our FR-DWBA analysis (equation $8$).

FIG. 4 : Same as FIG. 3 for the $2^{+}$\ ($16.922$) state of $^{8}$Be.

FIG. 5 : Same as FIG. 3 for the $1^{+}$\ ($17.640$) state of $^{8}$Be.

\end{document}